\documentclass[lettersize,journal]{IEEEtran}
\usepackage{amsmath,amsfonts,amssymb}
\usepackage{algorithmic}
\usepackage{algorithm}
\usepackage{array}
\usepackage[caption=false,font=normalsize,labelfont=sf,textfont=sf]{subfig}
\usepackage{textcomp}
\usepackage{stfloats}
\usepackage{url}
\usepackage{verbatim}
\usepackage{graphicx}
\usepackage{cite}
\usepackage{orcidlink}

\begin{document}
\bstctlcite{IEEEexample:BSTcontrol}

\title{The Kadomtsev pinch revisited for sheared-flow-stabilized Z-pinch modeling}

\author{D.W. Crews\orcidlink{0000-0003-0921-3515},~\IEEEmembership{Member, IEEE}
  I.A.M. Datta\orcidlink{0000-0002-0303-4017},
  E.T. Meier\orcidlink{0000-0001-6825-9842},
  and U. Shumlak\orcidlink{0000-0002-2918-5446},~\IEEEmembership{Fellow, IEEE}
  \thanks{D.W. Crews, I.A.M. Datta, E.T. Meier, and U. Shumlak, Theory and Modeling Division,~Zap Energy, Inc., Everett, WA}
  \thanks{U. Shumlak, Aerospace and Energetics Research Program,~University of Washington, Seattle, WA}
\thanks{Manuscript received October 10, 2023; revised January 22, 2024.}}

\markboth{
  IEEE TPS \copyright~2024 IEEE. Personal use is permitted, but republication/redistribution requires IEEE permission.}
{D.W. Crews, I.A.M. Datta, E.T. Meier, and U. Shumlak: Kadomtsev's Z pinch revisited}

\IEEEpubid{\copyright~2024 IEEE. This is the author's version which is not fully edited. Content has changed in the final version.  
  Citation information: DOI 10.1109/TPS.2024.3383312}

\maketitle

\begin{abstract}
  The Kadomtsev pinch, namely the Z-pinch profile marginally stable to interchange modes,
  is revisited in light of observations from axisymmetric MHD modeling
  of the FuZE sheared-flow-stabilized Z-pinch experiment.
  We show that Kadomtsev's stability criterion, cleanly derived by the minimum energy principle but of opaque physical significance,
  has an intuitive interpretation in the specific entropy analogous to the Schwarzschild-Ledoux criterion for
  convective stability of adiabatic pressure distributions in the fields of
  astrophysics, meteorology, and oceanography.
  By analogy, the Kadomtsev profile may be described as magnetoadiabatic
  in the sense that plasma pressure is polytropically related to area-averaged current density
  from the ideal MHD stability condition on the specific entropy.
  Further, the non-ideal stability condition of the entropy modes is shown to relate the specific entropy gradient
  to the ideal interchange stability function.
  Hence, the combined activity of the ideal interchange and non-ideal entropy modes
  drives both the specific entropy and specific magnetic flux gradients to zero in the marginally stable state.
  The physical properties of Kadomtsev's pinch are reviewed in detail
  and following from this the localization of pinch confinement, i.e., pinch size and inductance,
  is quantified by the ratio of extensive magnetic and thermal energies.
  In addition, results and analysis of axisymmetric MHD modeling of the FuZE Z-pinch experiment are presented
  where pinch structure is found to consist of a near-marginal flowing core surrounded by a super-magnetoadiabatic
  low-beta sheared flow.
  
\end{abstract}

\begin{IEEEkeywords}
  Z pinch, convective instability, interchange modes, entropy modes, sheared flow, self-organization
\end{IEEEkeywords}


\section{Introduction}
\IEEEPARstart{T}{he}
    self-field plasma confinement equilibrium, popularly known as the Z pinch, magnetically confines plasma pressure
    without external magnetic coils.
    Studies during the early days of plasma physics identified magnetohydrodynamic instabilities in pinches of axisymmetric
    and kink types, and the axisymmetric modes later fell under the broader category of interchange-type modes,
    including for example the tokamak ballooning modes.
    In fact, B.B.~Kadomtsev's seminal article ``Hydrodynamic stability of a plasma'' demonstrated the existence of a
    Z-pinch equilibrium marginally stable
    to the interchange modes by using the method of virtual displacement and the energy principle~\cite{kadomtsev}.
    For this reason, this equilibrium will be referred to here as ``Kadomtsev's pinch''.
    Kadomtsev's article noted that the interchange modes are analogous to buoyancy-driven modes in stratified fluids
    (e.g., Rayleigh-Taylor type),
    and expressed the stability condition in terms of the logarithmic pressure gradient as a rational function of
    plasma $\beta$ and adiabatic index $\gamma$.
    On the other hand, the same analysis concluded that the kink modes of azimuthal mode number $m=1$ were unstable
    for these interchange-stabilized distributions (or any other static distribution, for that matter).
    
    In the years since Kadomtsev's original work, the community's understanding of
    stratified buoyant fluids and magnetohydrodynamic
    interchange modes has considerably deepened.
    This deepening has centered around two somewhat complex elements,
    namely the role of entropy gradients and of sheared flows.
    In revisiting Kadomtsev's pinch, this work is intended to fill out the
    analogy between stable stratification and interchange stability,
    to clarify the role of specific entropy in defining the marginally stable state,
    and to shed light on the effect of sheared flow on interchange stability with evidence
    from axisymmetric resistive MHD simulations of Zap Energy's FuZE device.
    The intention is to reinterpret Kadomtsev's stability criterion
    from the somewhat nebulous, ``that the pressure does not fall off too rapidly with radius,''
    expressed by the logarithmic pressure gradient as a function of local plasma $\beta$ with
    an opaque physical significance,
    to an intuitive formulation in terms of the specific entropy and specific magnetic flux gradients.
    
    This article is organized as follows.
    Section~\ref{sec:lit_review} 
    reviews buoyant convection in fluid dynamics,
    the influence of sheared flows on interchange stability, and
    the Kadomtsev profile in the literature. 
    A theoretical exposition on Kadomtsev's pinch follows in Section~\ref{sec:properties}, 
    which first reviews the closely related phenomena of the adiabatic atmosphere and isentropic flows.
    Kadomtsev's pinch is studied in a physically motivated approach where
    the stability condition is expressed as a zero gradient of what some authors have
    termed magnetic entropy~\cite{pastukhov2000adiabatic}.
    The condition admits a first integral of the force balance which rearranges to Kadomtsev's implicit pressure profile,
    noted as analogous to meteorology's adiabatic pressure profile.
    Analytic properties are studied for the adiabatic indices $\gamma=2$, $\gamma=5/3$, and $\gamma\to 1$.
    The analysis is applied to the unity average beta concept in Section~\ref{subsec:macrobeta},
    and 
    utilized to quantify pinch radius as a function of the ratio of total magnetic-to-thermal energies.
    Lastly, Section~\ref{sec:sheared_flow} considers idealized theory for the flow velocity of a Z pinch downstream
    from a coaxial accelerator, and presents 2D axisymmetric simulations of the FuZE experiment
    demonstrating self-organized
    \IEEEpubidadjcol  
    flow pinch states where Kadomtsev's pinch profile plays a key role.
    
    \section{Interchange modes in the literature}\label{sec:lit_review}
    The hydrodynamic stability of stratified fluids is an important topic in all fluid dynamic disciplines with convective motions,
    including terrestrial and
    astrophysical meterology~\cite{defouw1970thermal},
    oceanography~\cite{riley2000fluid},
    stellar astrophysics~\cite{stein1974waves, anders2022schwarzschild},
    and plasma physics where convective instabilities arise as a category of ideal MHD instability
    known as interchange modes~\cite{kesner2003convective}.
    In the original summary of ideal MHD stability analysis through the energy method, Kadomtsev briefly observed
    that the physics of the interchange mode were analogous to that of general buoyant stratification~\cite{kadomtsev}.
    In Section~\ref{sec:properties} this analogy is recapitulated and the detailed mathematics worked out.
    
    On the other hand, the situation can first be summarized as follows.
    Like many fluid dynamical topics, convective instability is known as a complex phenomenon with a variety of control factors
    such as chemical composition~\cite{ledoux1947stellar}, sheared flow~\cite{case1960stability, kuo1963perturbations},
    radiation~\cite{loeb1997optical}, magnetization~\cite{gough1966influence}, external heating (Rayleigh-B\'{e}nard),
    discontinuous profiles (Rayleigh-Taylor), impulsive forcing by shock waves (Richtmyer-Meshkov), etc.
    The situation with flows is far from simple, since while the interchange is stabilized by sheared flow, the second derivative
    of transverse velocity is destabilizing (Kelvin-Helmholtz), and the question of stability under shear is rather complex due to
    the prevalance of nonmodal solutions in the initial-value problem~\cite{case1960stability_2},
    the related topic of stabilization of pseudomodes in the pseudospectrum~\cite{trefethen1993hydrodynamic},
    and three-dimensional perturbations~\cite{butler1992three}.
    Sheared flows, however, are generally understood to be stabilizing and to allow the persistance of
    superadiabatic profiles across the fluid dynamic disciplines~\cite{terry2000suppression}.
    Section~\ref{sec:sheared_flow} presents some simulation results and commentary on the role of sheared flows on
    the Kadomtsev Z-pinch profile.

    Regardless of the various complexities discussed above, fortunately the stability of a continuously stratified medium subject
    only to pressure and buoyancy is simple.
    The marginally stable profile in atmospheric meterology and astrophysics, known as the adiabatic atmosphere, is characterized by
    a vanishing gradient of the specific entropy $s$ such that $\nabla s\cdot\nabla p = 0$, a result known
    as Schwarzschild's criterion~\cite{schwarzschild2015structure}.
    This condition is widely used in astrophysics~\cite{thorne1966validity},
    metereology~\cite{mckenzie2011chandrasekhar}, and plasma physics~\cite{wahlberg2000stabilization, balbus2001convective}.
    The fundamental reason for the specific entropy to characterize the marginal state is that the displacement frequency due to
    buoyancy, known as the Brunt-V{\"a}is{\"a}l{\"a} frequency,
    is a function only of the invariants of motion along the displacement~\cite{haverkort2013magnetohydrodynamic}.
    As the conserved quantity in an adiabatic displacement is the specific entropy,
    the local buoyancy frequency is a direct measure of the specific entropy gradient.
    In Section~\ref{sec:z_pinch_adiabatic} we will see how Kadomtsev's pressure profile is precisely the analogous ``adiabatic'' profile
    for the interchange-stable pinch, which is characterized by the invariance of a quantity which
    Pastukhov~\cite{pastukhov2000adiabatic} and Kesner~\cite{kesner2003convective} refer to as magnetic entropy.
    It seems to the authors of this article that this entropy terminology does not correspond with that of
    Minardi~\cite{minardi2005magnetic}, which is properly developed from an information-theoretic perspective on
    the distribution of electric current.
    Therefore this article avoids the terminology of magnetic entropy for the defining invariant of Kadomtsev's pinch, which is instead
    referred to as the quantity $s_z$.
    
    The Kadomtsev-stable profiles make a frequent appearance in the Z-pinch literature.
    Kadomtsev profiles are observed as attractors in two-dimensional MHD simulations~\cite{self-org},
    and are utilized as initial conditions in simulations of sheared-flow stabilization~\cite{shumlak1995sheared, arber1996effect}.
    Although the Kadomtsev-stable profile is strictly unstable to the $m=1$ kink mode,
    under certain conditions it is observed experimentally.
    Chiefly these are the hardcore Z pinch and levitated dipole experiments which fix the magnetic axis with a conductor~\cite{kouznetsov2007effect}.
    Further, Kadomtsev's pinch has been invoked to explain data in compressional Z-pinch experiments~\cite{choi1988experimental, coppins1997review} and supporting theory~\cite{coppins1988ideal}.
    Finally, it is worth noting that also in astrophysics, meterology, and oceanography adiabatic profiles
    are often invoked alongside other
    organizing principles, and sometimes do not explain observations at all,
    due to the myriad of complications discussed above.
    For this reason the understanding of pressure, density, and temperature profiles in these disciplines
    remains an active research topic,
    as it certainly does in the field of magnetic plasma confinement.
    On the topic of self-organization in magnetic confinement with flows, the reader is pointed to work on double Beltrami fields
    and multiple-region relaxed magnetohydrodynamics~\cite{mahajan1998double, yoshida1999simultaneous, yoshida_beltrami_h_mode, dewar_yoshida_bhattacharjee_hudson_2015}.

\section{Properties of Kadomtsev's pinch}\label{sec:properties}
    
    \subsection{Analogy to stable stratification in the adiabatic atmosphere}\label{sec:adiabatic_atmosphere}
    Kadomtsev showed that the marginally stable pinch profile was implicitly a function of the local value of $\beta = p/p_B$
    with $p$ the thermal pressure and $p_B$ the magnetic pressure,
    \begin{equation}\label{eq:kadomtsev1}
      \frac{p_0}{p} = \Big(1 + \frac{\gamma-1}{\gamma}\frac{2}{\beta}\Big)^{\gamma/(\gamma-1)}.
    \end{equation}
    Note that $\beta$ is often defined in some integral manner, as discussed further in Section~\ref{subsec:macrobeta},
    yet this quantity is the local plasma $\beta$.
    Now, to make the analogy to stable stratification in the atmosphere clear,
    consider the static equilibrium of a stratified gas of mass density $\rho$ in a gravitational potential $\Phi_g$:
    \begin{equation}\label{eq:static_eql}
      -\nabla p - \rho\nabla\Phi_g = 0.
    \end{equation}
    Recall that the specific entropy $s$ of a perfect gas is given by $s = c_v\ln(p/\rho^\gamma)$ with $c_v$ the specific
    heat at constant volume and $\gamma$ the adiabatic index.
    Now noting that $(p/p_0) = (\rho/\rho_0)^\gamma e^{s/c_v}$ with $(p_0, \rho_0)$ reference constants making the
    logarithmic argument dimensionless, the pressure gradient is
    \begin{equation}
      \nabla p = c_s^2\nabla\rho + p \nabla s / c_v
    \end{equation}
    with $c_s^2 = \gamma p/\rho$ the local speed of sound squared.
    Substituting into Eq.~\ref{eq:static_eql} gives another form of equilibrium,
    \begin{equation}\label{eq:density_form_stratified}
      c_s^2 \nabla\rho = -\rho \nabla\Phi_g - (\gamma-1)\rho T \nabla s.
    \end{equation}
    Using the Schwarzschild condition $\nabla s = 0$
    means that the marginal state satisfies the equation
    \begin{equation}\label{eq:adiabatic_atmosphere_de}
      c_s^2\frac{d\rho}{dz} = -\rho \frac{d\Phi_g}{dz}.
    \end{equation}
    The solution to Eq.~\ref{eq:adiabatic_atmosphere_de} is called the adiabatic atmosphere.
    Note that in the isentropic solution the sound speed
    $c_s^2=c_s^2(r)$ is variable unless one takes the isothermal limit of $\gamma\to 1$.
    Expressed solely as a function of density for a particular adiabat, the sound speed is given by
    \begin{equation}
      c_s^2 = c_{s0}^2\Big(\frac{\rho}{\rho_0}\Big)^{\gamma-1}
    \end{equation}
    where $c_{s0}^2 = \gamma R T_0$. We note that specifying $\rho_0$ and $T_0$ is sufficient to specify the constant $e^{s_0/c_v}$
    defining the adiabat of the atmosphere, since any constant specific entropy $s_0$ satisfies the stability constraint.
    Defining $\widetilde{\rho} = \rho/\rho_0$, we obtain the separable differential equation
    \begin{equation}
      \widetilde{\rho}^{\gamma-2}\frac{d\widetilde{\rho}}{dz} = -c_{s0}^{-2}\frac{d\Phi_g}{dz}
    \end{equation}
    which with the initial condition $\rho(0) = \rho_0$ integrates into
    \begin{equation}\label{eq:adiabatic_atmosphere}
      \frac{\rho}{\rho_0} = \Big(1 - \frac{\gamma-1}{\gamma}\frac{m\Phi_g}{k_BT_0}\Big)^{1/(\gamma-1)}
    \end{equation}
    with the corresponding pressure solution $p = p_0(\rho/\rho_0)^\gamma$.
    It is clear that Kadomtsev's pinch profile given by Eq.~\ref{eq:kadomtsev1} is closely related to this pressure profile.
    Considering the isothermal limit $\gamma\to 1$ provides us with the physical meaning of this solution,
    \begin{equation}
      \rho = \rho_0\exp\Big(-\frac{m\Phi_g}{k_BT_0}\Big)
    \end{equation}
    as a Boltzmann distribution.
    For realistic values of $\gamma$ the isentropic solution is not isothermal, yet
    the marginal state is still nearly Boltzmann-distributed, albeit in the distribution's less
    familiar limiting form of $e^x=\lim_{n\to\infty}(1+x/n)^n$. 
    Therefore, in this work we refer to the form of Eq.~\ref{eq:adiabatic_atmosphere} as the frozen Boltzmann distribution.
    In the physical and mathematical literature distributions in the form of Eq.~\ref{eq:adiabatic_atmosphere} are referred
    to as $\kappa$- or $q$-Gaussian distributions and maximize a
    Tsallis-type entropy functional for the number density distribution~\cite{zhdankin_2023}.
    
    Finally, observe that Eq.~\ref{eq:adiabatic_atmosphere}, the adiabatic atmosphere, is a first integral of the momentum equation,
    and thus expresses constancy of specific enthalpy
    \begin{equation}\label{eq:conservation_of_enthalpy}
      c_pT + \Phi_g = c_pT_0,
    \end{equation}
    in the same way as the pressure-Mach number relation for compressible isentropic ($\nabla s=0$) flow
    \begin{equation}
      \frac{p_0}{p} = \Big(1 + \frac{\gamma-1}{\gamma}\frac{mv^2/2}{k_BT_0}\Big)^{\gamma/(\gamma-1)},
    \end{equation}
    is a rearranged form of a constant specific enthalpy 
    \begin{equation}
      c_pT + \frac{1}{2}v^2 = c_pT_0
    \end{equation}
    which similarly limits as $\gamma\to 1$ to a Boltzmann distribution in the kinetic energy.
    In these particular cases, the adiabatic equilibrium as a frozen Boltzmann distribution can be understood to consist of
    both constant energy and entropy per particle, completely determining the distribution provided that there are only two
    forms of energy.
    Yet the ideal MHD picture is more complex as there are three forms of energy.

\subsection{Pinch marginal stability in thermodynamic variables}\label{sec:z_pinch_adiabatic}
    In this section the marginal stability condition of the cylindrical plasma pinch to axisymmetric interchange
    modes is cast as a condition on the specific entropy and the specific magnetic flux gradients.
    This reformulation results in a particularly simple form of the stability criterion and of the marginally stable state,
    in particular in light of the activity of the entropy modes discussed in Section~\ref{subsec:entropy_modes}.
    The classical method to determine magnetohydrodynamic stability is accomplished analytically from
    the minimum energy principle provided that the force operator of a virtual displacement is self-adjoint~\cite{freidberg2014ideal}.
    According to this intuition, the reader is reminded of the equivalence in thermodynamics
    between the principle of minimum energy and the principle of maximum entropy~\cite{callen1998thermodynamics}.
    Now, it is well known that given such a virtual displacement of the static MHD plasma in equilibrium
    the force operator is self-adjoint.
    By this property, Kadomtsev showed that the energy of a virtual displacement of mode $m=0$
    was bounded in time provided that the plasma pressure gradient satisfied
    \begin{equation}\label{eq:kadomtsev_condition}
      -\frac{d\ln p}{d\ln r} \leq \frac{4\gamma}{2 + \gamma\beta}
    \end{equation}
    and that this is both necessary and sufficient for stability of the static Z-pinch~\cite{kadomtsev}.
    The condition may be recast in thermodynamic variables
    by considering the radial force balance
    \begin{equation}\label{eq:radial_force_balance}
      \frac{dp}{dr} = -\frac{B_\theta}{\mu_0r}\frac{d}{dr}(rB_\theta) = -\frac{1}{r^2}\frac{d}{dr}(r^2p_B).
    \end{equation}
    Next consider the thermodynamic identity $dh = v dp + Tds$, with $v$ specific volume,
    written in the form
    \begin{equation}\label{eq:first_law}
      \frac{dh}{dr} = \frac{1}{\rho}\frac{dp}{dr} + T\frac{ds}{dr}
    \end{equation}
    where $h = \frac{\gamma}{\gamma-1}\frac{p}{\rho}$ is the specific enthalpy and $s = c_v\ln(p/\rho^\gamma)$
    is the specific entropy with specific heat at constant volume $c_v = R/(\gamma-1)$.
    Note that in all of what follows, $T$ signifies the total plasma temperature $T = T_i + T_e$,
    since the distinction between the two temperatures is beyond the scope of ideal MHD.
    Equation~\ref{eq:first_law} is an algebraic identity between the change in enthalpy, the $pdv$ work,
    and the reversible non-$pdv$ work expressed as $T\nabla s$.
    In this sense as an identity, an analogous identity holds for the magnetic body force,
    \begin{equation}\label{eq:laplace_first_law}
      \frac{d}{dr}\Big(2\frac{p_B}{\rho}\Big) = \frac{1}{\rho r^2}\frac{d}{dr}(r^2p_B) +
      \frac{p_B}{\rho}\frac{d}{dr}\ln\Big(\frac{p_B}{\rho^2r^2}\Big).
    \end{equation}
    We recognize the specific magnetic enthalpy as $2p_B/\rho$ and the specific Laplace body force as $(\vec{j}\times\vec{B})_r/\rho$,
    leading us to label the quantities $T_m\equiv p_B/(\rho R)$ as a ``magnetic temperature'' and $s_m\equiv R\ln(p_B/(\rho^2r^2))$
    as a ``magnetic entropy''.
    In fact, $s_m$ is the specific magnetic flux (or frozen-in flux) of Alfven's law applied to the axisymmetric pinch,
    \begin{equation}
      \frac{d}{dt}\Big(\frac{B}{\rho r}\Big) = 0.
    \end{equation}
    It is distinct from both Pastukhov's notion of magnetic entropy in Ref.~\cite{pastukhov2000adiabatic}
    and Minardi's magnetic entropy
    concept based in information theory~\cite{minardi2001stationary}.
    However, it is clear that in magnetohydrodynamics the specific flux $B/\rho$ plays the role of an entropic coordinate
    in the sense of Eq.~\ref{eq:laplace_first_law}.
    Further, summation of Eqs.~\ref{eq:first_law} and~\ref{eq:laplace_first_law} gives the thermodynamic law for the
    magnetofluid,
    \begin{equation}\label{eq:mhd_thermodynamic_law}
      \frac{d}{dr}\big(c_pT + 2RT_m\big) - \frac{1}{\rho}\frac{d}{dr}\big(p + p_B\big) = T\frac{ds}{dr} +  T_m\frac{d}{dr}\Big(R\ln\big(\frac{p_B}{\rho^2}\big)\Big)
    \end{equation}
    as if ``$\gamma_m=2$'', a useful mnemonic.  
    It seems appropriate to point out here that the specific flux $B/\rho$ plays a double role as an entropic thermodynamic coordinate
    in the sense of Eq.~\ref{eq:mhd_thermodynamic_law}
    and as a vorticity in three-dimensional flow dynamics through the generalized (or canonical) vorticity~\cite{richardson2016effect}.
    
    Now with the variables $s$ and $s_m$, we eliminate pressure from Eq.~\ref{eq:radial_force_balance}
    in favor of the variables $(\rho, s, s_m)$,
    \begin{align}
      p &= \rho^\gamma e^{s/c_v}\\
      p_B &= \rho^2r^2 e^{s_m/R}
    \end{align}
    and after some algebra obtain a differential equation for $\rho$,
    \begin{equation}\label{eq:density_form}
      c^2\frac{d\rho}{dr} = -4\frac{p_B}{r} -\frac{1}{c_v}p\frac{ds}{dr} - p_B\frac{ds_m}{dr}
    \end{equation}
    where $c^2 = c_s^2 + c_a^2$ is the local magnetosonic speed.
    Equation~\ref{eq:density_form} is analogous to Eq.~\ref{eq:density_form_stratified}
    with the difference that the body force $\nabla\Phi_g$
    is due to the
    geometric term $4p_B/r$ of field line curvature and there are two thermodynamic variables $(s, s_m)$.
    Then dividing by magnetic pressure and replacing the density gradient by the gradient of pressure,
    we obtain a form of the static equilibrium as (with $c_p = \gamma c_v$),
    \begin{equation}\label{eq:kadomtsev_stability_alt}
      K(r) \equiv \frac{d\ln p}{d\ln r} + \frac{4\gamma}{2 + \gamma\beta} =
      \frac{2}{2 + \gamma\beta}\frac{d}{d\ln r}\Big(\frac{s}{c_p} - \frac{s_m}{2R}\Big)
    \end{equation}
    We recognize the left-hand side as Kadomtsev's stability function $K(r)$.
    Thus, the right-hand side is equivalent
    in equilibrium. Further, the logarithms combine to define $s_z$ as
    \begin{equation}\label{eq:entropy_combination}
      \frac{s_z}{R} \equiv \frac{s}{c_p} - \frac{s_m}{2R} = \ln\Big(\frac{p}{p_B^{\gamma/2}}r^{\gamma}\Big).
    \end{equation}
    so we write the marginal condition $K=0$ as
    \begin{equation}\label{eq:stable_pinch_stratification}
      \frac{ds_z}{dr} = 0.
    \end{equation}
    In this way, the Z-pinch is stably stratified to axisymmetric modes through Eq.~\ref{eq:stable_pinch_stratification}
    in a similar way as the stably stratified atmosphere,
    and the development of weak instability acts to bring the plasma towards this marginal state,
    as in the quasilinear theory of Ref.~\cite{kouznetsov2007quasilinear}.
    For example, for the axisymmetric-stable Bennett pinch of $\gamma=2$, the invariant $s_z$
    defines the Z-pinch adiabat through $\beta r^2 = e^{s_{z0}}$, and
    in the incompressible, isothermal limit $\gamma=1$, one has $\beta I_{\text{enc}} = e^{s_{z0}}$
    where $I_{\text{enc}}$ is the axial current enclosed at radius $r$.
    In general, though, this relationship can be expressed as the statement that 
    $p \sim \big(\frac{B}{r}\big)^\gamma \sim \big(\frac{I_{\text{enc}}}{\pi r^2}\big)^\gamma,$
    with equality given some constants which absorb into the adiabatic constant $e^{s_{z0}}$.
    The quantity $I_{\text{enc}} / \pi r^2\equiv \langle j_z\rangle$ is in fact the area-averaged current density,
    so the marginal state reflecting a vanishing gradient of $s_z$ has
    \begin{equation}\label{eq:polytropic_gamma}
      \frac{p}{p_0} = \Big(\frac{\langle j_z\rangle}{j_0}\Big)^\gamma
    \end{equation}
    which is evocative of the usual adiabatic condition $p = \rho^\gamma$,
    although the condition is nonlocal as the local value of pressure depends on the distribution of
    current within the plasma column up to that radius for $p=p(r)$.
    In analogy to meteorology, we refer to a pinch whose pressure exceeds the polytropic relation of Eq.~\ref{eq:polytropic_gamma}
    as super-magnetoadiabatic, in which case it is convectively unstable to magnetic flux interchange, and
    the opposite case as sub-magnetoadiabatic.
    In this terminology Kadomtsev's pinch is the magnetoadiabatic pinch.
    Just as in metereology, this terminology is simple but potentially misleading
    as the adiabatic atmosphere is not ``adiabatic,'' but rather isentropic, and
    the magnetoadiabatic pinch is also not ``adiabatic,'' but rather the specific magnetic flux $s_m$ is tied to
    the specific entropy $s$ such that $\nabla s_z=0$.
    
    In other words, Eq.~\ref{eq:entropy_combination} is a single equation in the two variables $s$ and $s_m$
    and thus any entropy profile $s=s(r)$ such that $\nabla s_m = 2(\gamma-1)/\gamma \nabla s$ represents a Kadomtsev pinch.
    This freedom is what is usually understood to be the freedom in the choice of temperature and
    density profiles for the equilibrium pinch.
    Two situations of note are the adiabatic pinch, where $\nabla s = 0$,
    and the isothermal pinch where $p / \rho = \text{const.}$
    In the former situation, the adiabatic pinch has both $\nabla s = 0$ and thus $\nabla s_m = 0$,
    so that density varies radially as $\rho \sim \langle j_z\rangle$.
    In the isothermal pinch, on the other hand, one finds that $\rho \sim \langle j_z\rangle^\gamma$.
    Both cases are magnetoadiabatic and therefore marginally stable to interchange.
    
    In the context of the fusion DT neutron yield, for temperatures up to approximately $5$ keV, $\dot{Y} \sim n^2T^4 \sim p^2 T^2$,
    such that for relatively low $T$ more concentrated temperatures are advantageous for the same pressure or supplied current,
    in which case the adiabatic pinch temperature profile appears to be advantageous.
    On the other hand, around $50$ keV the reaction rate becomes independent of temperature,
    making more concentrated density necessary for high-yield fusion conditions.
    Indeed, high densities are required for $Q>1$ conditions
    in a sheared-flow-stabilized Z pinch~\cite{shumlak2023fusion},
    in which case the isothermal profile is more desirable.
    However, nonideal instabilities known as entropy modes grow to relax $\nabla s\to 0$,
    making such instabilities important to model $Q>1$ Z pinches.

    \subsection{The entropy modes}\label{subsec:entropy_modes}
    Section~\ref{sec:z_pinch_adiabatic} found that the axisymmetric interchange instability drives the pinch
    toward a marginal state
    where the specific flux gradient $\nabla s_m$ is tied to the specific entropy gradient $\nabla s$
    through the interchange marginal stability condition.
    That the specific entropy is unconstrained, and consequently also the concentration of temperature, suggests that
    some critical physics may be missing from this picture of convective relaxation. 
    Kadomtsev in fact also determined this condition, building on work by a collaborator Y.A.~Tserkovnikov, to be a type of
    drift instability caused by the oppositely directed diamagnetic heat fluxes of electrons and ions in the
    inhomogeneous plasma pinch~\cite{kadomtsev1960convective}.
    
    Further, Kadomtsev's article anticipated that such drift modes, which leave thermal pressure unperturbed to first order with
    varying temperature and density, were in fact entropy fluctuations working to relax the temperature gradient,
    and hence are termed entropy modes.
    Unfortunately the dependence on $\gamma$ was suppressed in Kadomtsev's article in favor of the numerical factor of $5/3$.
    The important article on numerical entropy mode modeling by Angus et al.~\cite{angus_dorf_geyko} restored the $\gamma$-dependence,
    and gives the entropy mode stability condition
    \begin{equation}\label{eq:kadomtsev_stability_entropy_modes}
      \frac{d\ln T}{d\ln r} \leq 1 + \Big(\frac{2\gamma-1}{2\gamma} + \frac{\beta}{4}\Big)\frac{d\ln p}{d\ln r}.
    \end{equation}
    Noting that the specific entropy increment is given in terms of logarithmic temperature and pressure increments by
    \begin{equation}
      \frac{1}{R}ds = \frac{\gamma}{\gamma-1}d\ln T - d\ln p
    \end{equation}
    after some simple algebra one can show that Eq.~\ref{eq:kadomtsev_stability_entropy_modes} combined with
    Eq.~\ref{eq:kadomtsev_condition} simplifies to the condition
    \begin{equation}
      \frac{1}{c_p}\frac{ds}{d\ln r} \leq \frac{2+\gamma\beta}{4\gamma}K(r) \implies \frac{1}{c_v}\nabla s \leq \frac{1}{2}\nabla s_z.
    \end{equation}
    When relaxed to the marginally stable state the right-hand side is identically zero from
    Eq.~\ref{eq:kadomtsev_stability_alt},
    giving $\nabla s = 0$ for the specific entropy gradient.
    In summary, the Z-pinch entropy modes can be understood to relax the gradient of specific entropy, which combined
    with the interchange mode activity yields simple estimates for the fully relaxed profile as $\nabla s = \nabla s_m = 0$.

    \subsection{The Z-pinch buoyancy frequency}
    As pointed out in Refs.~\cite{kadomtsev1960convective,angus_dorf_geyko},
    the radial displacement frequency of
    a Z-pinch plasma parcel is found by linear analysis 
    assuming purely radial displacement to be
    \begin{equation}\label{eq:brunt_vaisala_zpinch}
      \omega^2 = \frac{2c_s^2}{r^2}K(r)
    \end{equation}
    with $K(r)$ the Kadomtsev stability function.
    It would then appear that the thermal sound speed is principally responsible for the growth of perturbations.
    However, using Eq.~\ref{eq:kadomtsev_stability_alt} for $\nabla s_z$, the sonic speed $c_s^2 = \gamma p/\rho$,
    Alfven speed $c_a^2=2p_B/\rho$, and magnetosonic speed $c^2 = c_s^2 + c_a^2$ gives a form more akin to the
    usual Brunt-V{\"a}is{\"a}l{\"a}
    frequency in cylindrical coordinates,
    \begin{equation}\label{eq:true_brunt}
      \omega^2 = \frac{2}{r}\frac{c_a^2c_s^2}{c^2}\frac{d}{dr}\Big(\frac{s_z}{R}\Big).
    \end{equation}
    Thus the speed appears as a combination of both characteristic velocities.
    To bring out the character of this quantity a bit, we can consider the case of an isothermal Bennett pinch with $\gamma=2$,
    where $c_a^2 = 2(r/r_p)^2\frac{kT}{m}$ and $c^2 = \gamma\frac{kT}{m}(1 + \frac{2}{\gamma\beta})$,
    \begin{equation}
      \omega_B^2 \approx \frac{r/r_p}{1 + (r/r_p)^2}c_s^2\frac{d}{dr}\Big(\frac{s_z}{R}\Big)
    \end{equation}
    such that the frequency of interchange is potentially greatest at the pinch radius $r=r_p$.
    Note that Eq.~\ref{eq:true_brunt} is not singular as $c_a^2$ always grows at least quadratically for center-peaked currents.

    \subsection{The marginal pressure distribution}
    Substituting the invariant $s_z$ into the pressure balance and integrating produces an energy relation,
    \begin{equation}\label{eq:energy_relation}
      \frac{\gamma}{\gamma-1}\Big(\frac{p}{p_0}\Big)^{(\gamma-1)/\gamma} + 2\Big(\frac{E_I}{P_0}\Big)^{1/2} = \frac{\gamma}{\gamma-1}
    \end{equation}
    with $E_I = \frac{\mu_0}{8\pi}I_{\text{enc}}^2$ and $P_0 = p_0 \pi r_p^2$.
    Each term represents fluid or magnetic specific enthalpy under adiabatic or flux-conserving conditions.
    That is, let the enthalpy of a plasma parcel isentropically
    brought to its reference pressure $p_0$ be the
    potential enthalpy $h_f^0 \equiv c_pT\big(\frac{p_0}{p}\big)^{(\gamma-1)/\gamma}$,
    and let the potential magnetic enthalpy
    $h_m^0 \equiv 2RT_m\big(\frac{P_0}{E_I}\big)^{1/2}$ be the magnetic enthalpy when brought to the
    reference magnetic pressure under constant specific flux.
    Then Eq.~\ref{eq:energy_relation} is
    \begin{equation}\label{eq:energy_integral}
      \frac{h_f}{h_f^0} + \frac{h_m}{h_m^0} = \frac{c_p}{R}.
    \end{equation}
    Equation~\ref{eq:energy_integral} has the significance that specific enthalpy is radially uniform for the adiabatic ($\nabla s=0$) pinch,
    and otherwise varies according to the particular entropy gradient of the profile.

    Now using the marginal stability condition as a differential equation between $\beta$ and $r$, we observe that
    $\beta E_I^{1/2} = (1 + 2(\gamma-1)/\gamma\beta)^{-1}$, so that Eq.~\ref{eq:energy_relation} rearranges to
    \begin{equation}\label{eq:kadomtsev_pressure}
      \frac{p_0}{p} = \Big(1 + \frac{\gamma-1}{\gamma}\frac{2}{\beta}\Big)^{\gamma/(\gamma-1)}.
    \end{equation}
    To reveal the physical character of the distribution, consider the limit $\gamma\to 1$ where the pressure varies as
    $p = p_0e^{-2/\beta}$, or 
    \begin{equation}\label{eq:boltzmann}
      p = p_0\exp\Big(-\frac{mv_a^2}{k_BT}\Big)
    \end{equation}
    where $v_a$ is the Alfven speed and $T$ is the total temperature.
    In other words, the pressure is a frozen Boltzmann distribution in terms of relative magnetic-to-thermal energy
    analogous to the isentropic relation for compressible flow.
    Then substituting Eq.~\ref{eq:kadomtsev_pressure} into the invariant $s_z$ one obtains
    \begin{equation}\label{eq:radius_relation}
      \Big(\frac{r}{r_p}\Big)^2 = \beta^{-1}\Big(1 + \frac{\gamma-1}{\gamma}\frac{2}{\beta}\Big)^{(2-\gamma)/(\gamma-1)}.
    \end{equation}
    This relation is an implicit relation for $\beta=\beta(r)$ in order to calculate the pressure,
    magnetic field, and current density profiles of the Kadomtsev-stable pinch.

    \subsection{Special cases of the marginal distribution}
\begin{figure}[t!]
      \centering
      \includegraphics[width=\linewidth]{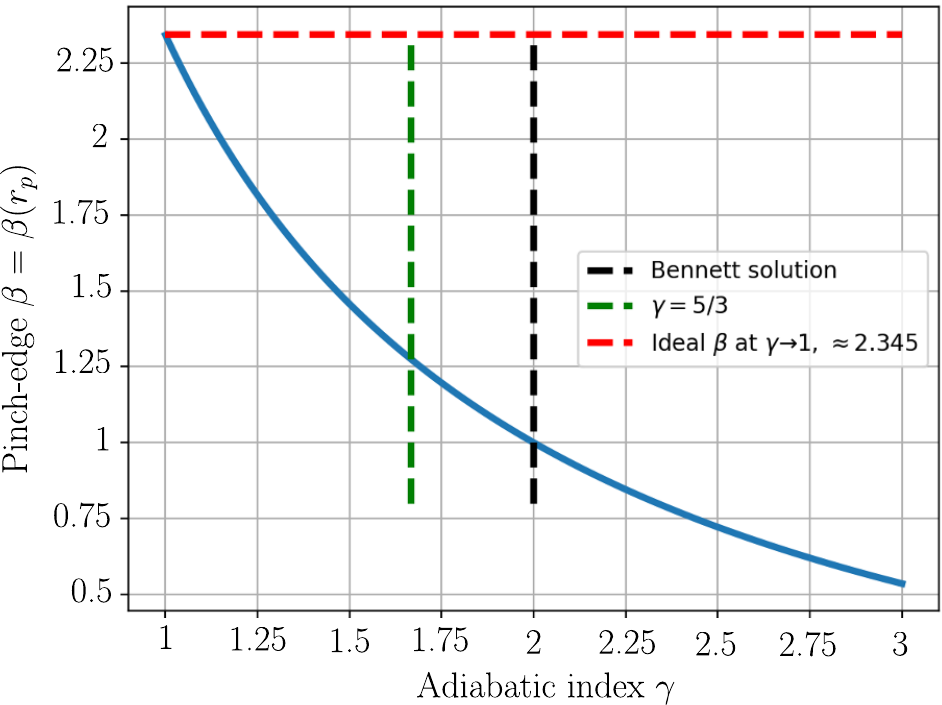}
      \caption{The local plasma $\beta=p/p_B$ at the characteristic radius $r_p$ (i.e., the pinch-edge beta)
        of the Kadomtsev-stable pinch increases as the adiabatic index $\gamma\to 1$.
        For the Bennett pinch, $\beta=1$ at the pinch radius, 
        and for $\gamma\in (1,2]$ the pinch-edge $\beta$ takes an intermediate value.}\label{fig:pinch_edge}
    \end{figure}

    A few interesting cases present themselves for Eqs.~\ref{eq:kadomtsev_pressure} and~\ref{eq:radius_relation}.
    The radius-$\beta$ relation is explicitly invertible for three cases of interest.
    Of course, the case $\gamma= 2$ obtains the Bennett profile where $(r_p/r)^2 = \beta$.
    There is also the limit $\gamma\to 1$ where $(r_p/r)^2 = \beta e^{-2/\beta}$ which inverts using the Lambert W function.
    In the particular case of $\gamma = 5/3$, Eq.~\ref{eq:radius_relation} is of the depressed cubic form
    \begin{equation}
      \beta^3 - \Big(\frac{r_p}{r}\Big)^4\beta - \frac{4}{5}\Big(\frac{r_p}{r}\Big)^4 = 0
    \end{equation}
    and thus possesses reasonably explicable roots for $\beta = \beta(r)$.
    A compact, explicit form particularly suitable for use in codes comes from Vi\`{e}te's form of the cubic root,
    \begin{equation}
      \beta = \Big(\frac{r_p}{r}\Big)^2\frac{2}{\sqrt{3}}\cos\Big(\frac{1}{3}\arccos\Big(\frac{6\sqrt{3}}{5}\Big(\frac{r}{r_p}\Big)^2\Big)\Big)
    \end{equation}
    considered as a complex-valued function returning a real-valued solution, which excludes the other complex-valued and negative-valued cubic roots.
    From these considerations, evidently local $\beta$ at the pinch edge radius depends on $\gamma$.
    For $\gamma=1$ at the pinch radius $r_p$, the solution to $2/\beta = e^{-1/\beta}$ is
    $\beta(r_p) = 2/W(2) \approx 2.34$
    where $W(r)$ is Lambert's W function.
    If the pinch radius were rescaled by $\sqrt{2}$ (as suggested by the relation $r_p^2 = \lambda_D^2 / 4 (c / v_e)$
    for the Bennett pinch~\cite{haines2011review}), then the pinch-edge beta satisfies
    $\beta(r_p / \sqrt{2})/2 = \Omega^{-1} \approx 1.763$ where $\Omega$ is the Omega constant of the Lambert W function.
    Transcendentals aside, evidently there is a tendency towards greater $\beta$ as $\gamma\to 1$, and this is accomplished at the expense
    of a larger specific heat for compression.
    Figure~\ref{fig:pinch_edge} shows the variation of pinch-edge radius $\beta$ with adiabatic index $\gamma$.
    
    In the infinite degree-of-freedom limit $\gamma\to 1$, the pressure has a simple explicit form
    in terms of the Lambert W function.
    With $2/\beta = W(2(r/r_p)^2)$, the Kadomtsev-stable pressure is
    \begin{equation}
      p(r) = p_0\exp\Big(-W(2(r/r_p)^2)\Big)
    \end{equation}
    and the profile for other values of $\gamma\in (1, 2]$ can be understood to limit between this form
    and the Bennett pinch
    whose explicit pressure profile is $p(r) = p_0/(1+(r/r_p)^2)^2$.
    From this, we note that the pinch-edge pressure is $p=p_0/4$ for the Bennett pinch
    and $p = W(2)/2 p_0 \approx 0.426p_0$ for
    the $\gamma\to 1$ limit.
    When pinch radius is rescaled by $\sqrt{2}$, then at $r=r_p/\sqrt{2}$ the pressure $p = \Omega p_0$
    and the magnetic pressure $p_B = \Omega^2p_0$.
    We also note another application of the invariant Eq.~\ref{eq:energy_relation} as an implicit relation
    for $p = p(r)$ valid for all $\gamma$ as
    \begin{equation}
      \frac{\gamma}{\gamma-1}\Big(\frac{p}{p_0}\Big)^{(\gamma-1)/\gamma} + 2\Big(\frac{r}{r_p}\Big)^{2}\Big(\frac{p}{p_0}\Big)^{1/\gamma} = \frac{\gamma}{\gamma-1}.
    \end{equation}

    \subsection{The pinch radius for marginally stable profiles}\label{subsec:macrobeta}
    In this section we demonstrate how the $\beta=\beta(r)$ relation expressing the distribution of the
    magnetic energy per thermal energy is closely connected to the general result for the inductance
    of a current-carrying pinch, and in this way we determine how the pinch's radius is related to its
    total thermal and magnetic energies.
    To begin, recall that Z pinches confine plasma with unity average $\beta$~\cite{shumlak2020z}.
    Precisely, the Bennett relation for self-magnetic confinement relates the self-magnetic energy
    of axial current  $\frac{\mu_0}{8\pi}I_{\text{enc}}^2$ to the excess thermal pressure contained within
    an area of cross-sectional radius $r$,
    \begin{equation}\label{eq:full_bennett}
      \frac{\mu_0}{8\pi}I_{\text{enc}}^2(r) = \pi r^2(\langle p\rangle - p(r))
    \end{equation}
    where $\langle p\rangle \equiv (\pi r^2)^{-1}\int_0^r p(r')2\pi r'dr'$ is the area-averaged thermal pressure.
    Out to a radius where $p\to 0$ Eq.~\ref{eq:full_bennett} gives
    \begin{equation}\label{eq:average_beta}
      \frac{\langle p\rangle}{p_B(r)} = 1
    \end{equation}
    with $p_B(r)$ the magnetic pressure at that radius.
    In this sense, the average Z-pinch $\beta$ is unity, and indeed
    non-zero axial magnetic field results in a value less than unity.
    Essentially, Eq.~\ref{eq:full_bennett} has the significance that purely axial current ideally confines
    an excess of thermal energy in an amount equal to
    the current's self-magnetic energy~\cite{shadowitz2012electromagnetic}.  

    Of course, in addition to the self-magnetic energy of the internal current,
    the total magnetic energy is fixed by the radius of the return current as in a coaxial cable.
    In the context of the internal conductor as a Z pinch, with fixed outer radius and variable inner radius, this is
    related to the spatial localization of the confined thermal energy (i.e., the internal conductor radius). 
    By combining the physics for the inductance of a coaxial conduit with the Bennett relation,
    we find that the Z-pinch radius is a function of a quantity
    for which both the numerator and denominator of Eq.~\ref{eq:average_beta} are averaged.
    
    To demonstrate, we analyze the sharp pinch for simplicity and then state results for
    some special cases of the Kadomtsev-stable profiles.
    Consider a sharp pinch of radius $r_p$ contained in a perfectly conducting wall of radius $r_w$
    returning all of the current passed by the sharp pinch.
    In this situation, the profiles of thermal and magnetic pressures are given by
    \begin{equation}
      p = \begin{cases} p_0, & r < r_p \\ 0, & r_p < r < r_w\\ 0 & r > r_w\end{cases},\quad
      p_B = \begin{cases} 0,& r < r_p\\ p_0\big(\frac{r_p}{r}\big)^2& r_p < r < r_w\\ 0 & r > r_w\end{cases}.
    \end{equation}
    Now 
    calculating the total thermal and magnetic energies per unit length
    by integrating over all radii we find a thermal energy per unit length in the amount of $P_T = p_0 \pi r_p^2$ and
    likewise magnetic energy per unit length of $P_B = p_0\pi r_p^2\ln\big(\big(\frac{r_w}{r_p}\big)^2\big)$.
    Now defining the ratio of linear pressure to magnetic pressure as
    $\beta_m \equiv P_T / P_B$ (i.e., Eq.~\ref{eq:average_beta} with both numerator and denominator area-averaged), one has
    \begin{equation}\label{eq:sharp_pinch_radius}
      \Big(\frac{r_p}{r_w}\Big)^2 = e^{-1/\beta_m}.
    \end{equation}
    Indeed, it is interesting to observe that
    the isothermal sharp pinch size is Boltzmann distributed as $\exp\big(-\frac{P_B/N}{k_BT}\big)$ with $N$ the linear density.
    The quantity $\beta_m$ measures the spatial concentration of pinch energy according to the usual physics of
    a coaxial conductor combined with the Bennett relation.
    Consequently, in order to localize the plasma in a small radius the parameter $\beta_m \ll 1$.
    For a fixed linear density $N = n_0\pi r_p^2$ with $n_0$ the volumetric particle density,
    high density requires a small pinch radius $r_p$,
    and high density pinch discharges are required for $Q>1$ conditions~\cite{shumlak2023fusion}.
    Therefore the total magnetic energy of a $Q>1$ Z-pinch discharge must be large compared to
    the confined thermal energy.
    However, we emphasize that this result occurs precisely because the Z pinch has unity average-$\beta$.
    That is, the local $\beta$ at the pinch edge is approximately unity,
    and most of the magnetic energy occuring in such a low-$\beta_m$ Z-pinch discharge
    is localized as either vacuum field or in a low-$\beta$ edge plasma.
    Indeed, for precisely this reason magnetically confined plasmas with axial magnetic fields
    such that $\beta$ is low everywhere consequently require even greater quantities of magnetic energy for thermal confinement.

\begin{figure}[t!]
      \centering
      \includegraphics[width=\linewidth]{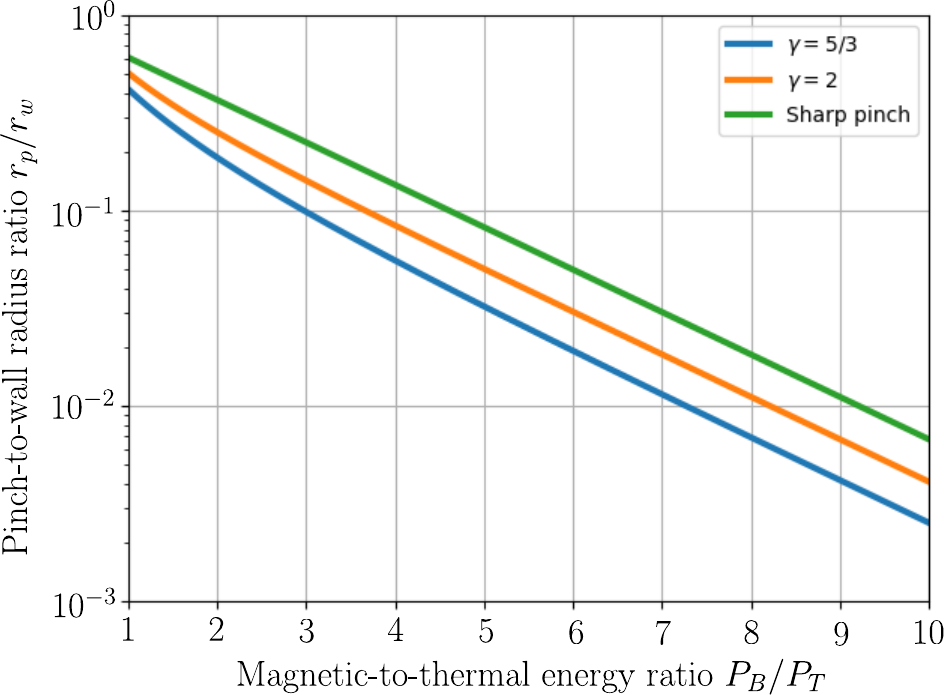}
      \caption{Z-pinch size asymptotically varies exponentially with the ratio of total confined thermal to magnetic pressures.
        Here $P_B$ is the magnetic energy per unit length and $P_T$ the pressure energy per unit length.
        For the Kadomtsev-stable profiles, smaller values of $\gamma$ give an increasing benefit over sharp pinches in terms of confinement,
        but at the expense of a lower temperature for the same confined pressure as the specific heat increases.}\label{fig:pinch_size}
    \end{figure}

    The result $r_p^2 \sim e^{-1/\beta_m}$ holds asymptotically for the Kadomtsev-stable Z-pinch profiles.
    Lacking a return current, the total magnetic energy is divergent for all values of $\gamma$.
    However, the magnetic energy is small enough for $r \gg r_p$
    that we may cut off the integral at the wall for a sufficient approximation.
    In the case of the Bennett pinch an exact solution can be found, but since the cut-off
    is not valid for $r_p \approx r_w$, we approximate $r_p\ll r_w$ for
    \begin{equation}
      \Big(\frac{r_p}{r_w}\Big)^2 \approx e^{-1}e^{-1/\beta_m}
    \end{equation}
    such that the diffuse Bennett pinch has a smaller radius than a sharp pinch of equivalent thermal and magnetic energies, being one $e$-folding factor smaller
    for the same value of $\beta_m$.
    A similar calculation of the Kadomtsev pinch for $\gamma=5/3$ demonstrates the equivalent asymptotic behavior as a function
    of $\beta_m$ with an additional benefit of approximately one more $e$-folding factor, as shown in Fig.~\ref{fig:pinch_size}.
    The physical reason for the asymptotic variation as $e^{-1/\beta_m}$
    is the inductive coupling of the current contained within $r\leq r_p$ and the return current
    at $r=r_w$, such that the inductance
    for $r_p \ll r_w$ is well-approximated by the usual coax factor $\ln(r_w/r_p)$ for $\beta_m \lesssim 1$.

\begin{figure}[t!]
      \centering
      \includegraphics[width=\linewidth]{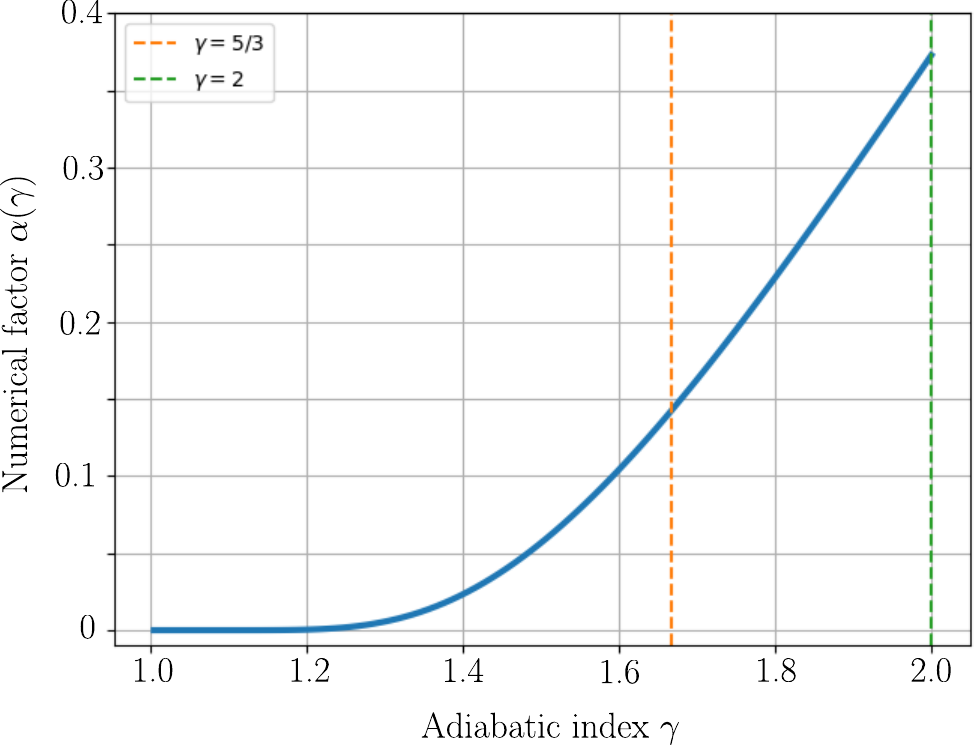}
      \caption{The functional dependence on adiabatic index for marginally Kadomtsev-stable pinch radius
        according to $(r_p/r_w)^2 = \alpha(\gamma)e^{-1/\beta_m}$ calculated numerically.
        The stable pinch radius collapses as $\gamma\to 1$ for the same magnetic and pressure energies $P_B, P_T$,
        or equivalently at fixed $\beta_m$.}\label{fig:pinch_size_fitting}
    \end{figure}

    For Kadomtsev's pinch, radius depends on adiabatic index $\gamma$ and the parameter $\beta_m$
    through $(r_p/r_w)^2 = \alpha(\gamma)e^{-1/\beta_m}$ (valid for $r_p\ll r_w$),
    where $\alpha(\gamma)$ is plotted in Fig.~\ref{fig:pinch_size_fitting}.
    Indeed, the total pressure energy per unit length of the Kadomtsev profile may be calculated as $P_T = p_0 \pi r_p^2 \big(\frac{\gamma}{2(\gamma-1)}\big)^2$.
    From this one can see that given fixed quantities of pressure energy $P_T$ and magnetic energy $P_B$, as $\gamma\to 1$ the pinch radius $r_p\to 0$ and the on-axis
    pressure $p_0\to \infty$ such that $p(r)\to P_T\delta(r)$.  

    \subsection{Inductance, enthalpy, and flux of Kadomtsev's pinches}\label{subsec:inductance}
    Section~\ref{subsec:macrobeta} introduced the energy measure $\beta_m \equiv P_T/P_B$ and its relationship to the
    characteristic pinch radius.
    Here the relation to inductance is made precise, and along the way a remarkable fact is noted that total pinch enthalpy
    $H = I_\infty \psi'$ where $\psi'$ is flux per unit length and $I_\infty$ total enclosed current.

    We begin by calculating the magnetic flux per unit length $\psi'=\int_0^r Bdr$ (or magnetic vector potential $A_z = -\psi'$) as
    \begin{equation}\label{eq:flux_per_unit_length}
      \frac{\psi'}{\psi'_0} = \frac{\gamma}{2\sqrt{2}(\gamma-1)^2}\Big(\log(1 + \Lambda) + \frac{\gamma-2}{1 + \Lambda^{-1}}\Big)
    \end{equation}
    from $\psi' = \int_\beta^\infty B(\beta')\frac{dr}{d\beta'}d\beta'$,
    where $\Lambda\equiv 2(\gamma-1)/(\gamma\beta)$ 
    and $\psi'_0 = B_0r_p$ with $B_0=\sqrt{\mu_0p_0}$ the characteristic flux density.
    Then computing the total enthalpy obtains (with $P_0 = p_0 \pi r_p^2$),
    \begin{equation}\label{eq:integral_enthalpy}
      H = \frac{\gamma}{\gamma-1}P_T + 2P_B = \sqrt{2}\frac{\gamma}{\gamma-1}P_0\frac{\psi'}{\psi'_0}.
    \end{equation}
    From $\mu_0 I_\infty^2/8\pi = (\gamma/(2(\gamma-1)))^2P_0$, we have
    $H = I_\infty\psi'$.
    Here $H$ and $\psi'$ are integrated up to a cut-off while $I_\infty$ is total current.
    Notably, under uniform axial flow $v_z$ enthalpy flux is precisely circuit power $Hv_z=\varphi I_\infty$ where voltage $\varphi = v_z\psi'$.
    Lastly, as $\psi' = L'I$ with $L'$ inductance per unit length, dividing Eq.~\ref{eq:integral_enthalpy} and applying the Bennett relation gives
    \begin{equation}\label{eq:inductance}
      L' = \frac{\mu_0}{4\pi}\frac{\gamma}{2(\gamma-1)}\Big(1 + \frac{2(\gamma-1)}{\gamma}\beta_m^{-1}\Big)\frac{\Lambda}{1+\Lambda}.
    \end{equation}
    Equation~\ref{eq:inductance} is well-approximated by $\Lambda/(1+\Lambda)\approx 1$ for cut-off $\beta\ll 1$. 
    For example, the $\gamma=2$ Bennett pinch inductance is $L' \approx \frac{\mu_0}{4\pi}(1 + \beta_m^{-1})$.
    Notable is the $\lim_{\gamma\to 1}L' = \frac{\mu_0}{4\pi}\beta^{-1}$.

    \section{The marginal profile and sheared axial flows}\label{sec:sheared_flow}
    This section begins with some theoretical remarks on sheared axial flows under completely ideal conditions
    in Z-pinch experiments,
    and then presents some observations from resistive MHD modeling of the FuZE experiment~\cite{shumlak2020z, levitt_zap}.
    In the sheared-flow-stabilized Z-pinch concept, a flow Z-pinch is formed downstream from a coaxial plasma gun.
    As a starting point, we draw some basic conclusions about steady flow using MHD theory in a similar
    manner to the analysis done by A.I.~Morozov and colleagues~\cite{morozov1980steady, morozov2012introduction}.
    To begin, recall that in compressible ideal MHD there are three axisymmetric streamline invariants, namely
    specific enthalpy, specific entropy, and specific magnetic flux:
    \begin{align}
      h_t \equiv& \frac{\gamma}{\gamma-1}\frac{p}{\rho} + 2\frac{p_B}{\rho} + \frac{v^2}{2},\\
      s \equiv& \frac{R}{\gamma-1}\ln\Big(\frac{p}{\rho^\gamma}\Big),\\
      s_m \equiv& R\ln\Big(\frac{p_B}{\rho^2r^2}\Big).
    \end{align}
    By expressing the convective derivative $(\vec{v}\cdot\nabla)\vec{v}$ in terms of its potential and vortex parts we may also formulate
    Crocco's theorem in steady state~\cite{liepmann2001elements},
    \begin{equation}
      \vec{v}\times\vec{\omega} = \nabla h_t - T\nabla s - T_m\nabla s_m
    \end{equation}
    where $\vec{\omega}=\nabla\times\vec{v}$.
    If we consider purely axial downstream flow such that flow forces are zero, the force balance is
    \begin{equation}\label{eq:force_balance_thermo}
      \frac{dh}{dr} = T\frac{ds}{dr} + T_m\frac{ds_m}{dr}
    \end{equation}
    where the specific MHD enthalpy $h = h_t - v^2/2$ does not include the kinetic energy.
    If the flow is self-organized into a near-Kadomtsev-stable profile with $\nabla s_z=0$, then 
    \begin{equation}\label{eq:enthalpy_gradient}
      \frac{dh}{dr} = \Theta\frac{ds}{dr}
    \end{equation}
    where $\Theta = T\big(\frac{p_0}{p}\big)^{(\gamma-1)/\gamma}$ is potential temperature (which is only constant radially
    for the adiabatic pinch with $\nabla s=0$).
    Equation~\ref{eq:enthalpy_gradient} follows by factoring $T\nabla s$ from Eq.~\ref{eq:force_balance_thermo}
    and eliminating $\beta=T/T_m$ using Eq.~\ref{eq:kadomtsev_pressure}.
    Expressing the specific entropy as $s = c_p\ln\Theta$ shows that Eq.~\ref{eq:enthalpy_gradient} integrates
    into the Kadomtsev pinch's specific enthalpy profile (which should be distinguished from the invariant of Eq.~\ref{eq:energy_integral}):
    \begin{equation}\label{eq:enthalpy_density}
      h = c_p\Theta = c_p T\Big(1 + \frac{\gamma-1}{\gamma}\frac{2}{\beta}\Big).
    \end{equation}
    As the specific enthalpy is conserved along streamlines, a first approximation is an isoenergetic condition
    transverse to the flow, \textit{i.e.}~$\frac{dh_t}{dr} = 0$, supposing that flow originates from one ``reservoir''.
    In the isoenergetic case the energy relation is
    \begin{equation}\label{eq:kinetic_energy_relation}
      c_p\Theta + \frac{v^2}{2} = \text{const.}
    \end{equation}
    Equation~\ref{eq:kinetic_energy_relation} means that the axial flow profile $v_z=v_z(r)$ of a pinch self-organized into
    a Kadomtsev-stable state downstream
    from its source is ideally a function only of the specific entropy profile $s=s(r)$ through $\Theta(r) = \Theta_0e^{s(r)/c_p}$.
    Because the pressure profile $p=p(r)$ is fixed by marginal stability,
    this is equivalent to thinking of the flow shear $\partial_r v_z$ as a function of the temperature profile $T=T(r)$
    in the first approximation.
    Further, Eq.~\ref{eq:kinetic_energy_relation} predicts that when $\nabla s = 0$ the axial flow is radially uniform.
    Because entropy modes drive $\nabla s\to 0$, this suggests that flow shear of a pinch marginally stable to both
    interchange and entropy modes must be controlled through engineering the gradient of specific enthalpy exiting the
    coaxial plasma accelerator, for example through an oblique shock wave attached to the tip of the central electrode.

    \begin{figure}[t!]
      \centering
      \includegraphics[width=\linewidth]{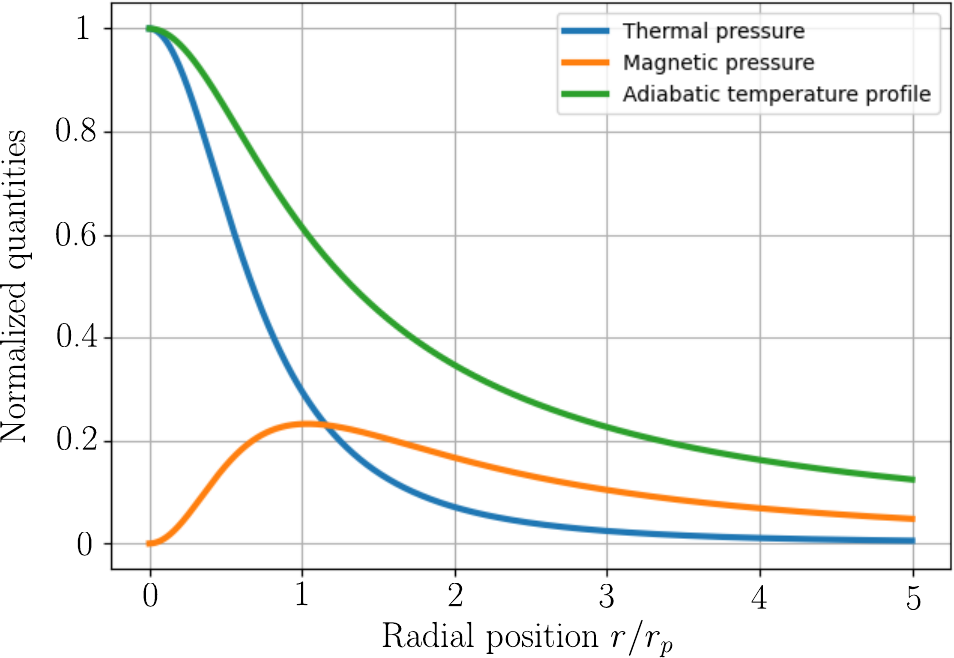}
      \caption{The thermal and magnetic pressures of the Kadomtsev pinch profile are depicted alongside the adiabatic
        temperature profile (where specific entropy $s$ is relaxed to $\nabla s=0$), with pressures normalized to $p_0$
        and the temperature normalized to $T_0$.
        The adiabatic temperature profile is marginally stable to the nonideal entropy modes for the pressure profile
        of Kadomtsev's pinch.
        Temperature profiles which decrease more slowly than the depicted, or even increase with radius,
        are superadiabatic ($\nabla s>0$) and unstable to the nonideal entropy modes
        similar to the Schwarzschild criterion for convective stability.}
      \label{fig:adiabatic_temperature}
    \end{figure}

    Considering isothermal pinch profiles, for a given uniform specific enthalpy $h_t$, say of the reservoir,
    Eq.~\ref{eq:kinetic_energy_relation} has no solution past some radius.
    For example, the Bennett pinch has $\beta = (r_p/r)^2$, such that the energy relation reads like
    $c_1 v^2 + c_2 r^2 = c_3$ for positive constants $c_1$, $c_2$, and $c_3$, defining an elliptical velocity profile, decreasing
    from a maximum on axis.
    This result occurs because isothermal pinches have specific enthalpy $h\to\infty$ as $r\to\infty$,
    which is problematic in experiment.
    Under the isoenergetic assumption, we can determine the conditions for axial velocity to be an increasing function of radius
    by writing Eq.~\ref{eq:kinetic_energy_relation} as
    \begin{equation}\label{eq:kinetic_energy_relation2}
      v^2 = 2h_{t0}\Big(1 - \frac{c_p\Theta_0}{h_{t0}}e^{s/c_p}\Big)
    \end{equation}
    with $h_{t0}$ the given constant specific enthalpy.
    Equation~\ref{eq:kinetic_energy_relation2} suggests 
    that for velocity to increase with radius, entropy should decrease with $\nabla s \leq 0$,
    which is also the entropy mode stability condition given marginal interchange stability.
    The temperature profiles with $\nabla s < 0$ are called sub-adiabatic.
    Such profiles are axially peaked and decrease at least as fast as the adiabatic profile
    $T(\beta) = T_0(1 + 2(\gamma-1)/\gamma\beta)^{-1}$ depicted in Fig.~\ref{fig:adiabatic_temperature}.
    These considerations suggest that adiabatic temperature profiles are to be expected rather
    than isothermal ones.
    On the other hand,
    velocity may freely increase with radius for arbitrary profiles of specific enthalpy.
    
    It must be cautioned here that our discussion is not self-consistent because the
    flow shear predicted by Eq.~\ref{eq:kinetic_energy_relation} for the marginally stable state
    also modifies the stability condition in a non-trivial manner,
    allowing for super-magnetoadiabatic pressure profiles~\cite{kouznetsov2007effect}
    (and most importantly, kink stability and three-dimensional self-organized states).
    There is also the issue of weak collisionality; the axisymmetric minimum energy state is Kadomtsev distributed
    only under the ideal fluid closure.
    The kinetic equilibrium of a weakly collisional Z pinch induces non-Maxwellian distribution functions
    in the presence of, \textit{e.g.}, axially sheared flow or temperature gradients, as the ion distribution function is distributed over the
    canonical momentum $P_z = mv_z + qA_z$ where $A_z$ is the magnetic vector potential~\cite{mahajan_1989}.
    This introduces intriguing phenomena such as a tendency for velocity to trend with the magnetic flux function~\cite{mahajan2000sheared, allanson2016}
    which are beyond the scope of the fluid model.
    That is to say, simulation of the appropriate model is necessary to self-consistently model flow pinch
    behavior in the weakly collisional regime of the FuZE experiment.
    

    \subsection{Results from MHD modeling of the FuZE experiment}\label{subsec:simulation_results}
\begin{figure}[t!]
      \centering
      \includegraphics[width=\linewidth]{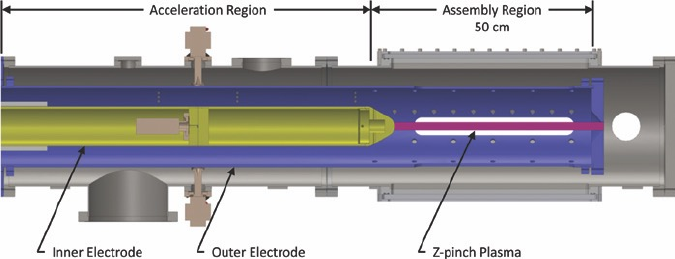}
      \caption{Diagram of the FuZE sheared-flow-stabilized Z-pinch device,
        showing the coaxial plasma accelerator of 100 cm length coupled to the 50 cm long pinch assembly region~\cite{shumlak2020z}.
        The phrase ``terminus of the inner electrode'' refers to the tip of the central conductor marking the
        end of the coaxial accelerator and the beginning of the pinch assembly region.}
      \label{fig:fuze_diagram}
    \end{figure}

    Here observations are presented of whole device
    two-dimensional axisymmetric resistive magnetohydrodynamic modeling of the
    FuZE experiment, a diagram of which is shown in Fig.~\ref{fig:fuze_diagram}, using the WARPXM discontinuous Galerkin
    finite element code~\cite{shumlak2011advanced}.
    The modeling approach employs an unstructured mesh of the whole device (including the accelerator)
    to solve the resistive MHD model with von Neumann--Richtmyer artificial viscosity coupled to
    a circuit model for the capacitor bank discharge.
    The simulations are initialized with plasma in the accelerator and the capacitor bank connected across the inner and outer electrodes.
    Simulations show plasma leaving the ``Acceleration Region'' (between the inner and outer electrodes)
    and forming a Z-pinch plasma in the ``Assembly Region'' (downstream from the terminus of the inner electrode).
    Related publications~\cite{iman_fuze, levitt_zap} include the whole device simulations used for the comparisons here.
    Please refer to Ref.~\cite{iman_fuze} for all details of simulation methodology.

    We investigate the role of marginal profiles in the data obtained by modeling.
    The results of this section are not meant to robustly model all aspects of the sheared-flow-stabilized Z pinch and its stability,
    in which three-dimensional, two-fluid~\cite{meier2021development}, and kinetic physics play key roles, but rather to demonstrate
    the robustness of attraction to organized states under rapid changes in the discharge, and to suggest the structure
    of the flow pinch to consist of a central, close-to-marginal flowing core surrounded by a low-$\beta$ sheared flow.
    We consider the profiles in two representative stages pre- and post- compression to examine the profiles at peak performance.
    
    Regarding the employed model, resistive MHD simulation certainly captures the drive toward marginal interchange stability. 
    However, only two-fluid modeling further captures the drive towards 
    the adiabatic pinch temperature profile through activity of the entropy modes.
    Further, as the pinches of the FuZE experiment are approaching the large Larmor radius regime,
    future research will consider the role of kinetic equilibrium and associated phenomena in the fluid description of weakly collisional flow Z pinches.

    \subsubsection{Diffuse flow pinch prior to compression, or ``zippering''}
    
\begin{figure}[t!]
      \centering
      \includegraphics[width=\linewidth]{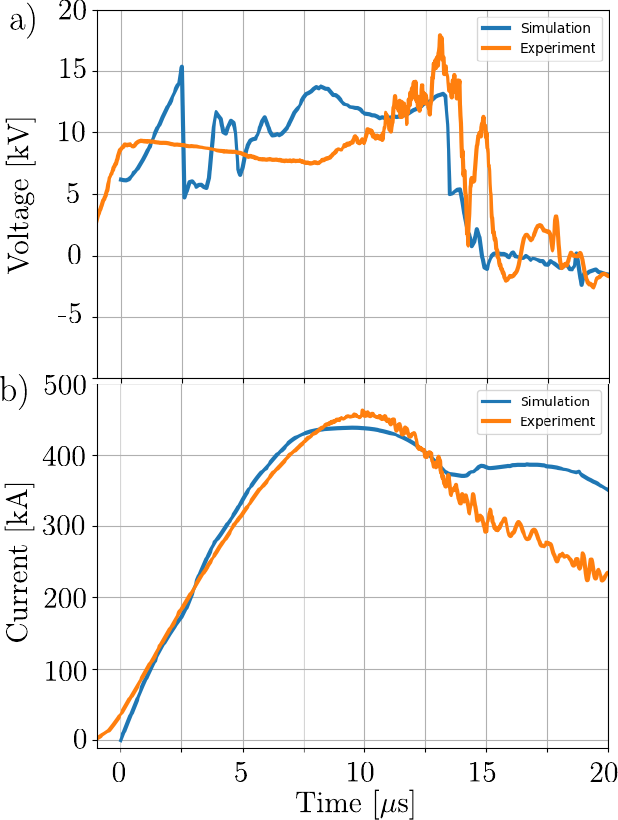}
      \caption{Current and voltage traces of a 25 kV FuZE discharge compared to the same traces from two-dimensional resistive MHD modeling.
        Specifically, this voltage is measured at the gap between the inner and outer electrodes.
        The simulation voltage is thought to overpredict conditions around $t=8$ $\mu$s due to a blow-by instability observed in simulation
        within the coaxial accelerator, which may not occur in this way under these experimental conditions, instead being modified by three-dimensional
        filamentation and the Hall effect.
        On the other hand, the peak voltage around $t=14$ $\mu$s is captured well.}
      \label{fig:kadomtsev_simulation_iv}
    \end{figure}

\begin{figure}[h!]
      \centering
      \includegraphics[width=\linewidth]{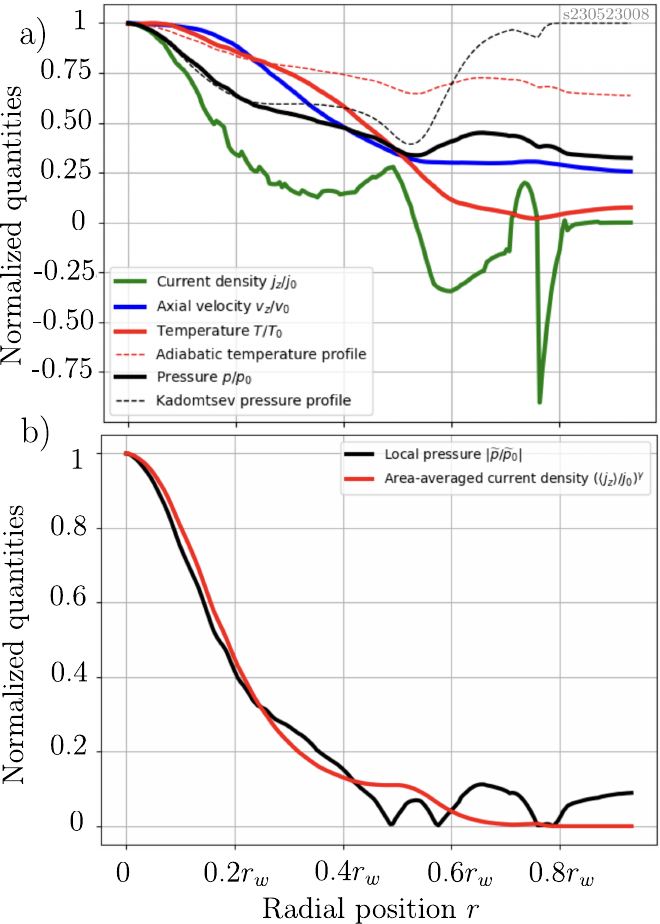}
      \caption{Simulated pinch radial profiles prior to ``zippering,'' at $t=8\mu$s in the discharge, measured $20$ cm downstream from the
        terminus of the coaxial discharge channel
        showing;
        a) radial profiles of current density, axial velocity, temperature, and pressure; and b)
        the corrected pressure $|\widetilde{p}|$ 
        and area-averaged current density $\langle j_z\rangle$ raised to the power of $\gamma$
        as $p \sim \langle j_z\rangle^\gamma$, \textit{i.e.}, the magnetoadiabatic profile.
        Corrected pressure is defined as $\widetilde{p} \equiv p - p_{\text{excess}}$ where $p_{\text{excess}}$ is the excess pressure
        taken to be $p_\text{excess}/p_0=0.38$, and must be used for magnetoadiabatic comparison because a uniform excess pressure
        is unknown to the MHD force balance.
        Here $r_w$ is the conducting wall of 10 cm radius.
        The Mach number at $r=0$ is $M\approx 1.8$.
        The dashed black line depicts the pressure profile if it were Kadomtsev-distributed with the local $\beta$, taking $\gamma=5/3$.
        The diffuse flow pinch approximately follows Kadomtsev's profile until a radius with significant
        shear where the distribution is super-magnetoadiabatic.
        Profile agreement ceases where the low-density edge plasma carries return current (where $j_z<0$).
        The flow pinch is observed to be stable for many Alfven transit times, hence the super-magnetoadiabatic profile
        around $r=0.3r_w$ could be attributed to linear flow shear.} 
      \label{fig:kadomtsev_simulation_1}
    \end{figure}

    Figure~\ref{fig:kadomtsev_simulation_iv} shows the current and voltage traces produced by experiment and simulation
    of a 
    typical high-performance FuZE discharge.
    Considering the flow pinch properties sufficiently far downstream from the terminus of the inner electrode
    of the coaxial accelerator such that the flow is primarily axial
    (specifically, $20$ cm downstream from the terminus),
    Fig.~\ref{fig:kadomtsev_simulation_1} shows the radial plasma profiles prior to pinch ``zippering''
    (also referred to as pinch assembly, as depicted in Fig.~3 of Ref.~\cite{shumlak_z_pinch}). 
    In this terminology, flow pinch zippering is similar to the classical zippering of
    gas puff pinch experiments~\cite{hussey1986large}.
    The Kadomtsev pressure profile of Fig.~\ref{fig:kadomtsev_simulation_1} is calculated by Eq.~\ref{eq:kadomtsev_pressure}
    and the local $\beta=p/p_B$ with $\gamma=5/3$. 
    
    It is thought that the flow pinch observed in Fig.~\ref{fig:kadomtsev_simulation_1} is only partially representative of
    experimental conditions due to the limitations of our model, but it is nevertheless interesting to observe
    the dynamics in the context of stability theory.
    To interpret the results, recall that linear flow shear (in ideal MHD, of either shear orientation) is stabilizing to the convective modes.
    The pinch is well described by a Kadomtsev-stable profile from the axis up to approximately the pinch radius, and
    axial velocity decreases with radius.
    In this situation the pinch core is primarily in a marginally stable equilibrium in which flow shear has not significantly altered
    the pressure distribution $p=p(\beta)$ away from Kadomtsev's pinch profile, and for which the flow shear is thought to 
    influence stability of the kink modes (or to encourage the existence of a three-dimensional relaxed state).
    The specific enthalpy is not radially uniform so that Eq.~\ref{eq:kinetic_energy_relation} does not directly apply,
    and the temperature and kinetic energy are observed to trend with one another.
    That $\nabla h_t\neq 0$ is thought to occur due an oblique shock upstream at the terminus of the central electrode
    which breaks the ideal streamline invariants including total specific enthalpy $h_t$.
    The thermal pressure profile tracks closely with the magnetoadiabatic expectation $p \sim \langle j_z\rangle^\gamma$ once
    corrected by an estimate for the excess background pressure which plays no role in the pinch equilibrium.

    \begin{figure}[h!]
      \centering
      \includegraphics[width=\linewidth]{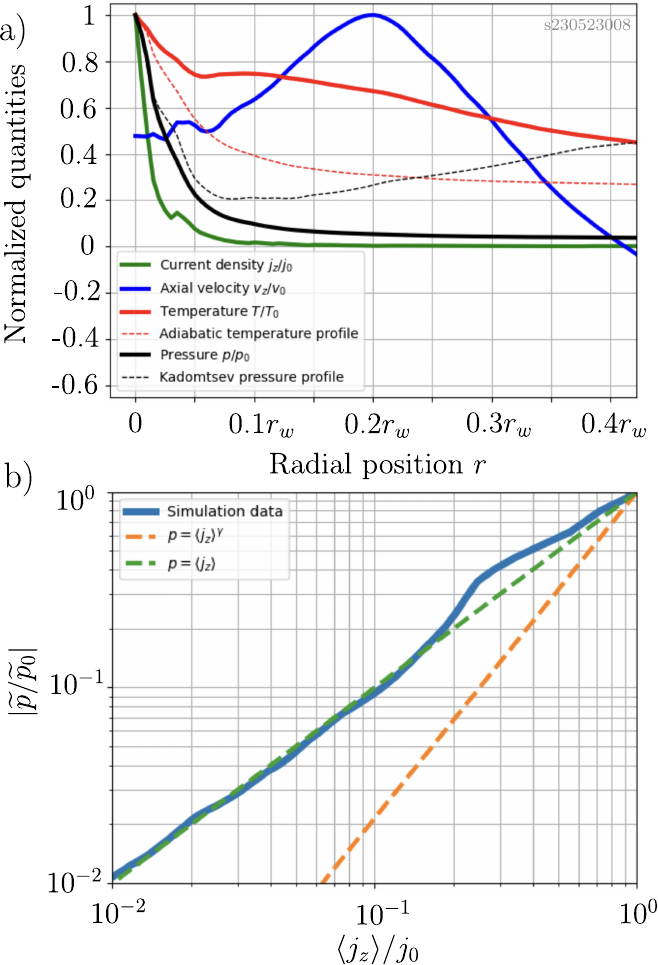}
      \caption{Conditions at time $t=13$ $\mu$s into the discharge,
        prior to the loss of flow power and just after ``zippering''
        into small radius and maximum performance
        (as measured by \emph{a posteriori} neutron yield rate calculation),
        with (a) the plot of
        Fig.~\ref{fig:kadomtsev_simulation_1}(a) zoomed radially up to $r=0.4r_w$,
        and (b) comparison of corrected pressure $|\widetilde{p}|$ to averaged current density $\langle j_z\rangle$ for the data of (a),
        where $p_\text{excess}=0.03 p_0$ (see Fig.~\ref{fig:kadomtsev_simulation_1} for definition of corrected pressure).
        In (b), the solid curve plots $\widetilde{p}=\widetilde{p}(\langle j_z\rangle)$ while the dashed lines are for reference.
        The profile is moderately sheared outside of the core region as the peak magnetosonic Mach number $M_\text{max}=0.2$.
        A super-magnetoadiabatic profile, i.e., with better-than-Kadomtsev confinement, is observed at all radii.
        Outside the pinch radius $r_p\approx 0.05 r_w$ the profile
        is observed to follow the super-magnetoadiabatic scaling $|\widetilde{p}| \sim \langle j_z\rangle$.
        This state is observed to last until the end of plasma voltage around $t=14$ $\mu$s in Fig.~\ref{fig:kadomtsev_simulation_iv},
        at which point the flow-sustaining electromagnetic power is extinguished.
        In the two-dimensional axisymmetric simulation the pinch then
        fully relaxes into a static Kadomtsev profile, yet in 3D would quickly undergo static kink instability.}
      \label{fig:kadomtsev_simulation_2}
    \end{figure}

    \subsubsection{Profiles of the post-``zippering'' compressed flow pinch}
    Following ``zippering'' of the pinch discharge,
    the diffuse flow pinch of Fig.~\ref{fig:kadomtsev_simulation_1}
    is near-adiabatically compressed when magnetic energy increases significantly,
    and the pinch core shrinks to a small radius in line with Section~\ref{subsec:macrobeta}.
    Figure~\ref{fig:kadomtsev_simulation_2} shows such a simulated post-zippering flow pinch
    and a comparison of its pressure to area-averaged current density. 
    The profile within the compressed pinch radius is close to marginal 
    in the sense that expected $m=0$ growth times consist of many magnetosonic transit times, as described below,
    and the profile is surrounded by a low-$\beta$ edge plasma with sheared axial flow deviating from the marginal state.
    A layer of edge plasma around $r=0.5r_w$, out of view of the figure, carries some return current.
    A super-magnetoadiabatic profile is observed in the low-$\beta$ plasma layer in-between the pinch edge
    and the return current-carrying edge.
    Namely, an ``isothermal-like'' profile $p \sim \langle j_z\rangle$ (better than Kadomtsev) is observed in this layer
    rather than the magnetoadiabatic profile $p \sim \langle j_z\rangle^\gamma$.
    A clear interpretation of this observation is difficult,
    but we do note that every polytropic relation $p \sim \langle j_z\rangle^a$
    corresponds with some polytropic MHD equilibrium.

    To clarify the phrase ``close to marginal'' used above, consider that if the profile were strongly unstable we could
    estimate from the displacement frequency of Eq.~\ref{eq:brunt_vaisala_zpinch}
    a growth-time $\tau\approx r_p/c$ where $c$ measures the magnetosonic speed
    (with $K\approx 1$ at the pinch edge for a strongly unstable profile).
    Using $r_p\approx 0.5$ cm and $c\approx 6.5\times 10^5$ m/s we find a growth-time of $\tau \approx 10$ ns, which is not observed.
    However, using the observed scaling $p \sim \langle j_z\rangle$, we estimate $\tau^{-2}\approx 2(\gamma-1)\frac{c^2}{r_p}(-\nabla\ln p)$ from Eq.~\ref{eq:true_brunt}
    and obtain $\tau\approx 3.5$ $\mu$s, 
    which is comparable to the steady flow-through and dynamical timescales.
    
    In addition to the single simulation examined here, we have considered simulations conducted under a variety of alternative realizations
    of discharge voltage, current, and capacitor bank configuration.
    Conditions similar to the results discussed previously are observed in each realization.
    Namely, pinch pressure and current are, prior to ``zippering'', Kadomtsev-distributed (\textit{i.e.}, following the polytropic relation $p \sim \langle j_z\rangle^\gamma$)
    with $\beta_m \lesssim 1$.
    The pressure distribution breaks from marginal stability following compression, yet is often observed to remain polytropic with an exponent less than $\gamma$.
    We leave a thorough investigation of this phenomenon to future work.
    
    To summarize, we observe that axisymmetric compression of the marginal profile remains near-marginal and benefits from enhanced stability.
    As the compressed profile is close to the static marginal profile it is likely undergoing slow $m=0$ relaxation on the order of $1-10$ $\mu$s
    (with an indeterminable role played by flow shear), but voltage decreases on a faster time-scale.
    We point the reader to Ref.~\cite{iman_fuze} for an in-depth discussion of the simulation's macroevolution. 
    

    \subsubsection{Observed temperature profiles}
    The core plasma temperature and low-$\beta$ edge plasma temperature
    are observed to be super-adiabatic, meaning that temperatures drop off more slowly than if
    specific entropy were radially constant.
    We expect that with higher fidelity physics entropy mode activity would relax temperature to the adiabatic profile.
    While two-fluid and three-dimensional physics are expected to change the nature of the marginal state attained in modeling,
    it seems likely that this conceptual picture of a core plasma with relaxed gradients surrounded by low-$\beta$ sheared flow
    will continue to hold under appropriate conditions, in line with the stepped relaxation region approach
    used with success in MHD modeling~\cite{dewar2008relaxed}.
    We expect three-dimensionality to complicate this picture in suprising ways,
    for example the saturated state attained by weakly unstable kink modes in a sheared-flow-stabilized Z pinch
    with conducting wall boundary.

    \section{Conclusion}
    This work revisited the Kadomtsev profile describing axisymmetric marginal stability of the Z pinch
    and investigated its thermodynamic properties,
    filling a gap in the literature for the static profile.
    Self-organization based on thermodynamic principles in the magnetic confinement and plasma turbulence
    communities is a fruitful topic of
    research~\cite{hasegawa1985self, minardi2001stationary, dewar2008relaxed, hasegawa2014fusion},
    so it is hoped that this article contributes to further research in this field by elucidating the properties
    of self-organization in one of its simplest solvable manifestations other than the force-free Taylor states,
    namely the axisymetrically constrained Z pinch.
    In this simple model, we combine the well-known result from the minimum energy principle together
    with the thermodynamics of a perfect gas to demonstrate that the state marginal to both interchange
    and drift-entropy modes is one where the gradients of both specific entropy and magnetic flux are zero.
    This result is happily in accordance with the equivalence between minimum energy and maximum entropy principles,
    and establishes a clear analogy to the Schwarzschild-Ledoux stability criterion used in other branches of hydrodynamics.
    However, we emphasize that the core result, namely Kadomtsev's stability function, is most easily determined by Kadomtsev's
    minimum energy method.
    We do point out, however, that an interesting avenue for future research is to deduce equilibrium stability as a condition
    on $s_z$ using an entropy functional method.
    
    In addition, the analogy between Kadomtsev's pinch and the
    adiabatic profile used in meteorology and oceanography
    was employed to draw out parallels to sheared-flow stabilization of the respective interchange modes in these fields,
    and how these marginal states are closely related to statistically probable distributions, here referred to as frozen
    Boltzmann distributions.
    Indeed, Kadomtsev's pinch is precisely this sort of frozen Boltzmann distribution in the magnetic energy.
    This was then extended to a demonstration in Section~\ref{subsec:macrobeta} that the pinch radius itself
    also follows an exponential relation in the ratio of magnetic to thermal energies.
    It was also observed in Section~\ref{subsec:inductance} that the extensive total enthalpy $H$ of the Kadomtsev pinch
    scales with magnetic flux per unit length $\psi'$ as $H = I_\infty \psi'$ for all values of $\gamma$,
    with the significance that unstable profiles like the sharp pinch can be said to contain an excess ``unmixed'' thermal enthalpy.
    
    Further, theory is presented here suggesting that non-isothermal temperature profiles are more likely
    than isothermal ones in the collisional fluid regime according to the activity of the entropy modes.
    This conclusion is further supported by the results of
    Section~\ref{sec:sheared_flow} where it was shown that the specific enthalpy of an
    isothermal Kadomtsev-stable pinch diverges as $\beta\to 0$ in the edge plasma,
    but the specific enthalpy is constant for adiabatic (i.e., constant specific entropy) profiles.
    Indeed, the simulation results of the FuZE experiment shown in Section~\ref{subsec:simulation_results} demonstrate
    non-isothermal profiles even without two-fluid physics.
    These simulation results indicate attraction to the marginal state in the flow pinch formed downstream
    from a coaxial accelerator, and the sustainment of a near-marginal state following pinch ``zippering''/compression.

    \section*{Acknowledgments}
    \noindent The authors would like to thank the entire team at Zap Energy for their efforts in producing and analyzing data
    from the FuZE and FuZE-Q experiments.
    We would also like to thank the Theory and Modeling group, including Jonny Dadras, Noah Reddell,
    Steve Richardson, Christine Roark, Peter Stoltz, and Whitney Thomas
    for their efforts in whole device modeling of the
    sheared-flow-stabilized Z-pinch experiments.
    We would especially like to thank Anton Stepanov and Derek Sutherland for thought-provoking discussions
    at the 2023 International Dense Z-Pinch conference,
    and in particular Anton Stepanov for discussions regarding Section~\ref{subsec:macrobeta}.

    The information, data, or work presented herein was funded in part by the Advanced Research Projects Agency –
    Energy (ARPA-E), U.S. Department of Energy, under Award Nos. DE-AR-0000571, DE-AR-0001010, DE-AR-0001260 and
    by the Air Force Office of Scientific Research under Grant No. FA9550-15-1-0271,
    and is also based in part upon work supported by the National Science Foundation under Grant No. PHY-2108419.
    This research used resources of the National Energy Research Scientific Computing Center (NERSC),
    a U.S. Department of Energy Office of Science User Facility located at Lawrence Berkeley National Laboratory,
    operated under Contract No. DE-AC02-05CH11231.

    \bibliography{main}
    \bibliographystyle{IEEEtran}

\vfill

\end{document}